\documentclass[aps,prl,superscriptaddress,twocolumn]{revtex4-2}
\usepackage{amsmath}
\usepackage{graphicx}	
\usepackage{bm} 
\usepackage{color}
\usepackage{setspace}
\usepackage{placeins}
\usepackage{natbib}
\usepackage{ulem}
\begin{document}
\title{Entanglement-enhanced correlation propagation in the one-dimensional SU($N$) Fermi-Hubbard model}
\date{\today}
\author{Mathias~Mikkelsen}
\email[]{mathias@qunasys.com}
\affiliation{QunaSys Europe, Copenhagen, Denmark}
\author{Ippei~Danshita}
\email[]{danshita@phys.kindai.ac.jp}
\affiliation{Department of Physics, Kindai University, Higashi-Osaka, Osaka 577-8502, Japan}
\begin{abstract} 
We investigate the dynamics of correlation propagation in the one-dimensional Fermi-Hubbard model with SU($N$) symmetry when the repulsive-interaction strength is quenched from a large value, at which the ground state is a Mott-insulator with $1/N$ filling, to an intermediate value. From approximate analytical insights based on a simple model that captures the essential physics of the doublon excitations, we show that entanglement in the initial state leads to collective enhancement of the propagation velocity  $v_{\text{SU}(N)}$ when $N>2$, becoming equal to the velocity of the Bose-Hubbard model in the large-$N$ limit. 
These results are supported by numerical calculations of the density-density correlation in the quench dynamics for $N=2,3,4,$ and $6$. 
\end{abstract}

\maketitle

{\it Introduction.} Exquisite controllability of quantum simulators, such as optical lattices loaded with ultracold gases~\cite{Gross,Schafer}, Rydberg atom arrays~\cite{Browaeys,Morgado}, trapped ions~\cite{Blatt,Wu,Monroe}, and superconducting circuits~\cite{Lamata,He}, have provided unique opportunities for studying non-equilibrium dynamics of closed quantum systems. In particular, experiments simulating the single-component Bose-Hubbard model (SCBHM) with nearest-neighbor hopping~\cite{Cheneau,Takasu} and $S=1/2$ spin models with interactions that algebraically decay with distance $r$ as $\sim r^{-\alpha}$~\cite{Richerme, Jurcevic, Chen} have addressed dynamical spreading of correlations. The question of how quantum correlations propagate in isolated quantum systems which undergo unitary time evolution is a fundamental problem having attracted interest for decades in the context of quantum-information propagation and thermalization~\cite{Gong,Cheneau,Cheneau-2}. Previous theoretical studies have shown that when the interactions are short-ranged or the power $\alpha$ of the algebraic interactions is sufficiently large, the correlations exhibit light-cone-like propagation with a certain speed~\cite{Barmettler,Natu,Hauke,Fitzpatrick,Cevolani,Nagao-1,Despres,Nagao-2,Mokhtari-Jazi,Schneider,Kaneko,Nagao-3}. 

In Refs.~\cite{Cheneau,Takasu,Richerme,Jurcevic,Chen}, the observation of light-cone-like propagation was indeed reported. In these examples, the initial state is a simple product state, namely a Mott insulator with unit filling or a spin-polarized state, such that the correlation propagations are qualitatively understood using the quasi-particle picture, in which quasi-particles excited on top of the product state carry correlations. Such advances in understanding quantum correlation propagations naturally pose the question of whether there are other intriguing mechanisms of correlation propagation starting from more complicated entangled initial states.

Quantum simulators of the Fermi-Hubbard model (FHM) with SU($N$) symmetry, which are composed of $N$-flavor Fermi gases of alkali-earth(-like) atoms in optical lattices~\cite{Taie-1,Hofrichter,Ozawa,Taie-2,Tusi,Pasqualetti}, have been remarkably developed recently~\cite{Schafer,Cazalilla,He-2,Padilla}. Thanks to the Pomeranchuk cooling mechanism, the $1/N$-filled Mott insulating state of the one-dimensional (1D) system with $N=6$ has been cooled down to $T \simeq 0.1 J/k_{\rm B}$ in recent experiments~\cite{Taie-2}, where $J$ is the hopping energy. At such a low temperature, the system is expected to exhibit the behavior of a $(N-1)$-component Tomonaga-Luttinger liquid~\cite{Bonnes}, which is highly entangled in comparison with the Mott insulator of the SCBHM. The entanglement entropy is proportional to $N-1$~\cite{Capponi} so it can be systematically controlled by varying $N$. Moreover, quantum-gas microscope techniques with single-site resolution have been recently applied to alkali-earth(-like) atoms in optical lattices~\cite{Yamamoto,Miranda,Buob}, allowing for access to the dynamics of a density-density correlation in the SU($N$) FHM. 

In this paper, we analyze the correlation-propagation dynamics of the 1D SU($N$) FHM after a quantum quench starting from a $1/N$-filled Mott insulator. From approximate analytical calculations based on the assumption that the correlations are carried by doublon-holon excitations, we find that the effective hopping of doublon excitations increases with $N$ due to high entanglement of the particle-excitation states, and that it coincides with the hopping of particle excitations in the SCBHM in the limit of $N\rightarrow\infty$. 
Furthermore, using matrix product state (MPS) methods~\cite{Schollwock}, we numerically calculate the time evolution of the density-density correlation function to show that the correlation-propagation velocity indeed increases with $N$ due to the enhanced hopping.

{\it Model and protocol.}
We consider the 1D SU($N$) FHM,
\begin{eqnarray}
\hat{H}_{{\rm SU(}\!N\!{\rm )}} &=& -J \sum_{\alpha=1}^N\sum_{j=1}^{L-1} \left(\hat{c}_{\alpha,j}^\dagger \hat{c}_{\alpha,j+1}+{\rm H.c.}\right)
\nonumber \\
&&+\, U \sum_{\alpha \neq \beta }\sum_{j=1}^{L} \hat{n}_{\alpha,j} \hat{n}_{\beta,j},\label{eq:SUNHubbard}
\end{eqnarray} 
where $\hat{c}_{\alpha}$ is the annihilation operator of flavor-$\alpha$ fermion at site $j$, $\hat{n}_{\alpha,j}\equiv \hat{c}_{\alpha,j}^\dagger\hat{c}_{\alpha,j}$,  $J$ the hopping energy, $U(>0)$ the onsite repulsive interaction, and $L$ the number of sites.
This model quantitatively describes $N$-flavor 1D Fermi gases with SU($N$) symmetry confined in optical lattices~\cite{Taie-1,Hofrichter,Ozawa,Taie-2,Tusi,Pasqualetti}. 
We are interested in propagation of quantum correlations captured by the time evolution of the density-density correlation function,
\begin{align}
D(r,t)= \langle\psi(t)| (\hat{n}_{j+r}-\bar{n}) (\hat{n}_j -\bar{n}) |\psi(t)\rangle,
\label{eq:Densitycorrelations}
\end{align}
where $\bar{n}\equiv N_{\rm P}/L$ is the average number of particles per site, $N_{\rm P}$ the total number of particles, and $\hat{n}_j=\sum_\alpha \hat{n}_{\alpha,j}$.
$D(r,t)$ can be measured in experiments using quantum-gas microscopy techniques~\cite{Cheneau}. As an initial state, we consider the ground state of Eq.~(\ref{eq:SUNHubbard}) with $\bar{n}=1$ in the limit of $U/J\rightarrow \infty$, which is a Mott insulating state. At $t=0$, we set $U/J$ to be finite and calculate the unitary time evolution, $|\psi(t)\rangle = e^{-i\hat{H}t/\hbar}|\psi(0)\rangle$, with the time-independent Hamiltonian. In short, we analyze dynamics subjected to a quantum quench from $U/J\rightarrow \infty$ to a finite value of $U/J$. Hereafter, we set $\hbar = 1$ and $a=1$ except in the figures, where $a$ is the lattice spacing.

For the SCBHM, the Mott insulating ground state at $U/J\rightarrow \infty$ is a simple product state with no entanglement, namely, $\bigotimes_{j=1}^L |\bar{n}\rangle_j$, where $|\bar{n}\rangle_j$ is the local Fock state at site $j$. In the present case, by contrast, the Mott insulating ground state has separated charge and flavor excitation sectors~\cite{Capponi}. While the charge excitation is gapped due to the relevant Umklapp processes, the flavor sectors form a $(N-1)$-component Tomonaga-Luttinger liquid. The gapless flavor excitation branches contribute to half-chain entanglement entropy of the system, which is approximately proportional to $N-1$. We will see that such relatively large entanglement plays a crucial role in the correlation-propagation dynamics.

{\it Analytical calculations.} Let us analytically obtain an approximate expression of the correlation-propagation velocity. For the SCBHM, when the initial state is an MI state with $\bar{n}=1$, the correlation is carried dominantly by a hole-doublon excitation~\cite{Cheneau,Barmettler}. We expect this to also be the case for the SU($N$) FHM so we focus on how a holon and a doublon propagate in the medium of the $\bar{n}=1$ MI state. While a doublon is always accompanied by a holon in the full model, considering the two separately allows us to understand their properties better.

\begin{figure*}
\centering
\includegraphics[width=1\linewidth]{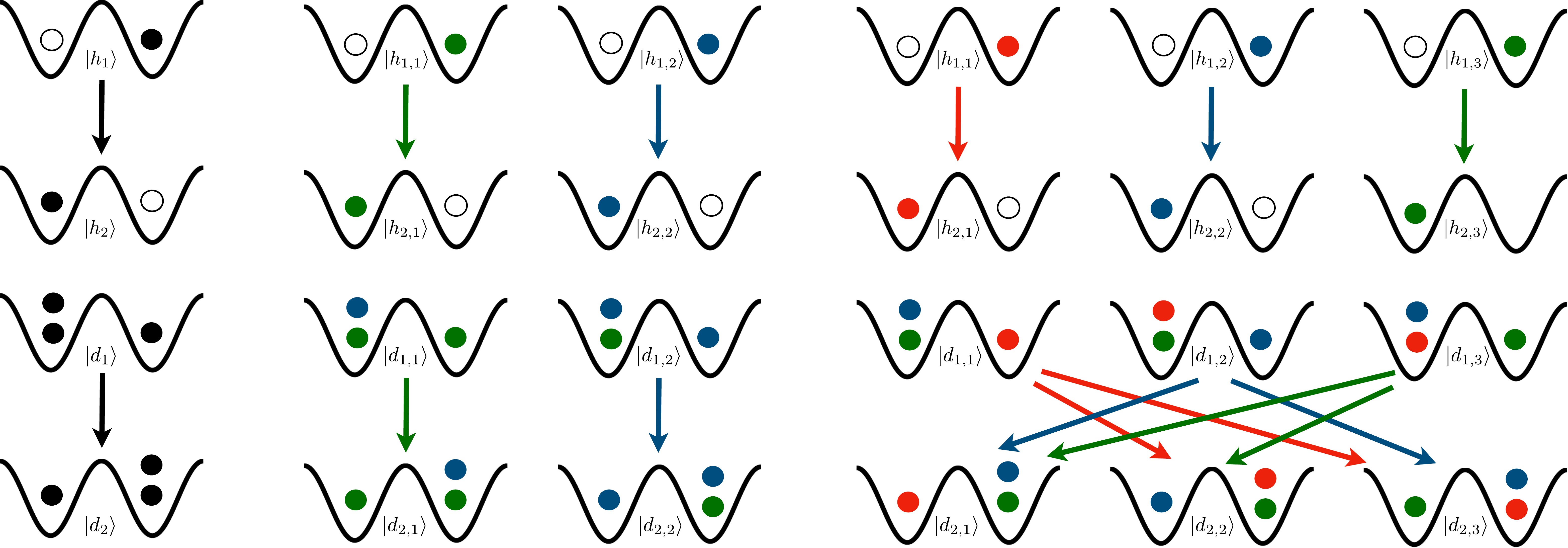}
\caption{Illustration of a simple two-site model for SCBHM, SU(2) and SU(3) FHM (for simplicity we restrict SU(3) to the ground-state symmetry sector $N_{\rm A}=N_{\rm B}=N_{\rm C}=1$). The green, blue and red circles correspond to different flavors, while the empty circle corresponds to a hole (an empty site). We can define the SU(2) and SU(3) equal-superposition collective doublon and holon states as $|d_j \rangle = \frac{1}{\sqrt{N}}\sum_{\nu=1}^{N} |d_{j,\nu}\rangle$ and $|h_j \rangle = \frac{1}{\sqrt{N}}\sum_{\nu=1}^{N} |h_{j,\nu}\rangle$, where $j=1,2$.}
\label{fig:twositemodel} 
\end{figure*}

To illustrate the important physics, we first consider a simple 2-site model for SU(2) and SU(3). On the basis of this, one can obtain a rough estimate of the velocity at which an excitation can move between two adjacent sites. In Fig.~\ref{fig:twositemodel} the holon $|h_j\rangle$ and doublon $|d_j\rangle$ states and connections between possible configurations for the SCBHM, SU(2), and SU(3) FHMs are displayed. 
These connections are possible in the sense that they are connected by the hopping term of the Hamiltonian, each with a matrix element corresponding to $\pm J$. For the SCBHM, it is well-known that the  propagation speed of the holon and doublon are qualitatively different due to bosonic enhancement. In the two-site model, this is observed in the matrix elements for connecting the two sites as $\langle h_1 |\hat{H}_{\text{BH}}| h_2 \rangle = -J $ and  $\langle d_1 |\hat{H}_{\text{BH}}| d_2 \rangle = -2J$. This factor comes from the definition of the bosonic creation and annihilation operators. For the SU(2) case, no such enhancement is possible as doublons correspond to two different flavor fermions and each doublon configuration on the left only connects to one doublon configuration on the right. Indeed, in terms of the collective state defined in Fig.~\ref{fig:twositemodel}, $\langle h_1|\hat{H}_{\text{SU(2)}}| h_2 \rangle = -J $ and  $\langle d_1|\hat{H}_{\text{SU(2)}}| d_2 \rangle = -J$. For the SU(3) case, however, 
due to the presence of multiple doublons any configuration with a doublon on the left will connect to two configurations on the right and vice versa. Conversely any configuration with a holon on the left side will only connect to one holon configuration on the right side.  This means that the matrix elements for the collective states defined in Fig.~\ref{fig:twositemodel} are equivalent to the SCBHM, that is $\langle h_1 |\hat{H}_{\text{SU(3)}}| h_2 \rangle = -J $ and  $\langle d_1 |\hat{H}_{\text{SU(3)}}| d_2 \rangle = -2J$. This is a crucial observation which also holds true for larger $N$ and larger $L$.

In order to understand this more systematically, we consider the following two models: (i) $N_{\rm F}$ particles of each flavor, with $N_{\rm P}=N N_{\rm F}$ total particles in $L=N_{\rm P}-1$ sites, restricted to the Fock space containing one double occupation with the remaining sites having unit filling. (ii) $N_{\rm F}$ particles of each flavor, with $N_{\rm P}=N N_{\rm F}$ total particles in $L=N_P+1$ sites, restricted to the space containing one hole with the remaining sites having unit filling. For $N=2$, this does not correspond to the situation in Fig.~\ref{fig:twositemodel} in the sense that the minimal multi-site models consistent with the SU(2) case is $N_{\rm P}=4$ and $L=3,5$, although the physics is very similar. As we give a detailed explanation of these models in the supplemental material \cite{supplement}, we only summarize some of the most important features here. One can always define a model in terms of an effective lattice with $L$ sites and a large local Hilbert space of dimension $Q_{\rm d}$ or $Q_{\rm h}$, where  
$Q_{\rm d}=\frac{N(N-1) (N_{\rm P}-2)!}{2\left((N_{\rm F}-1)!\right)^2\left(N_{\rm F}!\right)^{N-2}}$ or $Q_{\rm h}=\frac{N_{\rm P}!}{(N_{\rm F}!)^N}$ is the number of states that have one doublon or holon at a given physical site. The equal superposition of all such local states $|d_j \rangle = \frac{1}{\sqrt{Q_{\rm d}}}\sum_{\nu=1}^{Q_{\rm d}} |d_{j,\nu}\rangle$ and $|h_j \rangle = \frac{1}{\sqrt{Q_{\rm h}}}\sum_{\nu=1}^{Q_{\rm h}} |h_{j,\nu}\rangle$ are useful for understanding the physics of these models. 

The behavior for holons observed in the simple two-site model holds for any $N$, that is, they will only ever connect to one adjacent holon state. The equal-superposition collective states $|h_j \rangle$ form a disconnected sector in the Hamiltonian with matrix-elements that correspond to a standard single-particle Hamiltonian with hopping $J$. For periodic boundary conditions this holon sector forms the energy band  $E(k)=-2J \cos(k)$ so that the maximum group velocity of the holon is $2J$, where $k=\frac{2\pi n}{L}$ with $n=0,...,L-1$. There will be no sectors with a larger bandwidth indicating that there is no enhancement for the holon velocity as expected. 

As we saw in the simple two-site example, $N>2$ leads to doublon configurations that connect to two states with a doublon on the adjacent site. For $N_{\rm P}>N$, however, not all the terms in the collective state connect to two states. One can use combinatorial arguments to count the number of states $Q_1$ or $Q_2$ that connect to one or two adjacent states respectively~\cite{supplement}, which determines the matrix-element between two adjacent superposition states  
 $\langle d_j | \hat{H}_{\text{SU(}\!N\!{\rm )}}|d_{j+1} \rangle=-K$ as
\begin{align}
\frac{K}{J}=\frac{2Q_2+Q_1}{Q_1+Q_2}= \frac{N-1-\frac{1}{N_{\rm F}}}{\frac{N}{2}-\frac{1}{N_{\rm F}}}.
\label{eq:collectivehoppingK}
\end{align}

For $N=2$, where $Q_2=0$, $K=1$ at all values $N_{\rm F}$. In this case, similar arguments as for the holon case lead to a disconnected sector in terms of the superposition states with energies $E(k)=-2J \cos(k)$. Hence, the SU(2) case has no enhancement for the doublon velocity. For $N=N_{\rm P}$ (the illustration of Fig.~\ref{fig:twositemodel} corresponds to $N=N_{\rm P}=3$), $Q_1=0$ and $K=2J$. The equal-superposition collective state  will therefore always have matrix elements $\langle d_j | \hat{H}_{\text{SU(}\!N\!{\rm )}}|d_{j+1} \rangle=-2J$. Similarly to the previous cases, the matrix-elements of these collective states form a disconnected sector in the Hamiltonian governed by a free particle Hamiltonian, but now with an effective hopping of $2J$. For periodic boundary conditions the energies are therefore given by $-4J \cos(k)$ so that the maximum group velocity of the doublon is $4J$.

The ground-state sector in the large $N$ limit of $N=N_{\rm P}$ is fully equivalent to results obtained from the SCBHM doublon model. In fact, this equivalence is not only observed for these simple excitation models, but is a general feature for the ground state sector of SU($N$) fermionic Hamiltonians when $N=N_{\rm P}$. In the supplemental material \cite{supplement}, we show that states obeying SU($N$) symmetry such as the ground state are described by a disconnected subsector equivalent to single-component bosonic Hamiltonians with the same parameters.

For $N>2$ and $N_{\rm P}>N$, the equal doublon superposition states will not separate into their own sector and obtaining analytical results for the band structure, which is closely related to the propagation velocity, is difficult. However, the value $K$ obtained as the matrix element between the naive equal superposition states gives some indication of the propagation speed. In the supplemental material, we show results of exact diagonalization calculations of the actual spectrum~\cite{supplement}. These show that a naive calculation based on $K$ slightly underestimates the bandwidth, because not all the states in $Q_1$ and $Q_2$ will be in an equal superposition in the relevant linear combinations that determine the minimum and maximum energy. 
Nonetheless, the results suggest that the behavior of the bandwidth is qualitatively similar. $K$ asymptotically decreases to a value $\lim_{N_{\rm P}\rightarrow \infty} K = 2(1-1/N)J$ as the system size increases. This asymptotic value itself grows asymptotically with $N$ towards the SCBHM equivalent value of $K=2J$ in the limit of $N \rightarrow \infty$. These results strongly suggest that velocity enhancement will be observed for $N>2$ even when $N_{\rm P} \gg N$ and that the enhancement qualitatively behaves like $2(1-1/N)$.

\begin{figure*}
\centering
\includegraphics[width=1.0\linewidth]{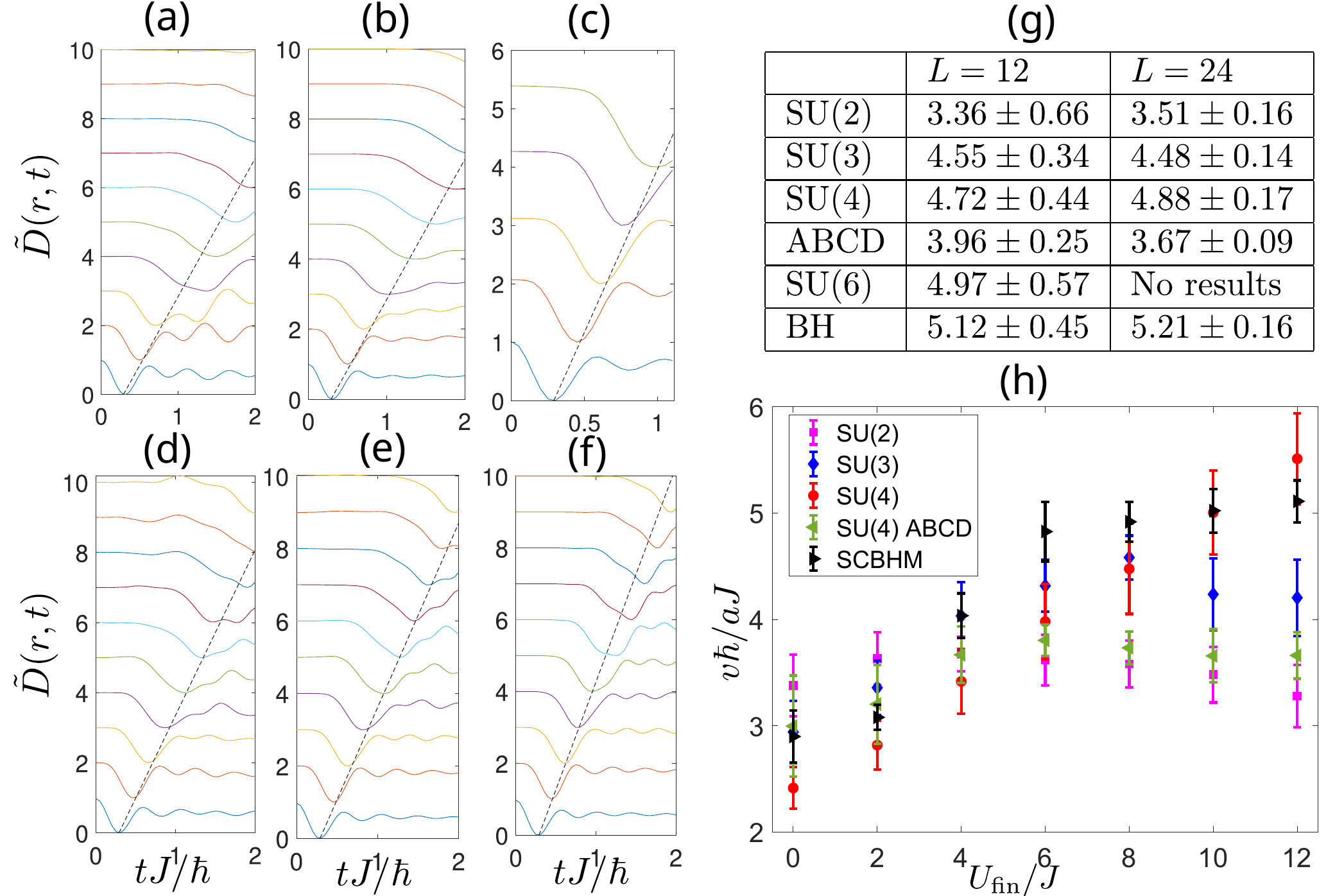}
\caption{ (a)-(f) shows $\tilde{D}(r,t)$ of systems time-evolved with a final Hamiltonian that has $U_{\rm fin}/J=8$ for several distances $r$. The curves for different $r$ are shifted for clarity by $r$. (a),(c-f) shows this for initial states which are ground states of $U_{\rm ini}/J=40$ for the respective Hamiltonian types SU(2) (a), SU(6) (c), SU(3) (d), SU(4) FHM (e) and SCBHM (f) , while (b) shows the ABCD initial state time-evolved by the SU(4) Hamiltonian. For SU(6), we consider $L=12$, while $L=24$ in all the other cases. The dashed lines represents $v(t-t_1)$, where $t_1$ corresponds to the location of the first dip in each plot and $v$ is given by $v_{\text{SU}(N)}=\left(2K+2J\right)a/\hbar$, $v_{\text{ABCD}}=4Ja/\hbar$ and $v_{\rm SCBHM}=6Ja/\hbar$ (g) shows the velocities $v \hbar/ aJ$ extracted by a fitting for $U_{\rm ini}/J=40$ and $U_{\rm fin}/J=8$ when $L=12$ and $24$. (h) shows fitted velocities $v\hbar/ aJ$ for different values of $U_{\rm fin}/J$ when $L=24,U_{\rm ini}/J=40$ (the ABCD case has no value $U_{\rm ini}/J$).  The magenta squares correspond to SU(2), green left-facing triangles to ABCD, the blue diamonds to SU(3), the red circles to SU(4) and the black right-facing triangles to the SCBHM.} 
\label{fig:FullModelResults} 
\end{figure*}

{\it Numerical calculations.} Having explained the underlying physical mechanism of the enhancement of the propagation velocity, 
we investigate the dynamics of the full SU($N$) FHM numerically with use of the time evolving block decimation method~\cite{Vidal} based on the MPS representation of quantum many-body states~\cite{Schollwock} (See also Supplementary material~\cite{supplement} and Ref.~\cite{Mikkelsen} for specific MPS representations in the present case). The expected enhancement is only possible for superpositions of doublon states. This means that in terms of the original model, the initial state must have a large degree of entanglement in the flavor sector corresponding to superpositions of different flavors on each site. The existence of enhancement is therefore strongly dependent on the initial state and can only be observed for quantum superposition states, which means that high temperature will destroy the effect. Similarly a classical product state with no entanglement in the flavor sector should lead to no enhancement. We investigate the SCBHM and SU($N$) FHM with $N=2, 3, 4,$ and $6$ using the ground state at $U_{\rm ini}=40J$, which we obtain by means of the density matrix renormalization group (DMRG) method~\cite{Schollwock,White}, as our initial state. 
Additionally we investigate a product state $|{\rm ABCD}....{\rm ABCD}\rangle$ time-evolved by the SU(4) FHM to confirm that enhancement is not present in the absence of entanglement. In order to more easily observe the propagation of correlations, it is useful to normalize $D(r)$ for each value $r$ with the maximum absolute value it takes during time evolution, i.e., 
\begin{eqnarray}
\tilde{D}(r,t)= \frac{D(r,t)}{\text{Max}_{t\in [0,T]}\left(D(r,t)\right)}.
\label{eq:DensitycorrelationsNormalized}
\end{eqnarray}

Our main focus is on the case $U_{\rm fin}/J=8$, where we expect the enhancement but the interaction is still small enough that accurate numerics can be obtained. In Figs.~\ref{fig:FullModelResults}(a)-(f), the time evolution of $\tilde{D}(r,t)$ for several $r$ is plotted for the SU($N$) FHM with $N=2, 3, 4,$ and $6$, and the SCBHM (see the figure caption for more details). In all cases, we see a characteristic dip in $\tilde{D}(r,t)$ for fixed distances $r$. The time at which the dip forms a minimum changes linearly with $r$, indicating a light-cone-like propagation of the correlation~\cite{Cheneau,Barmettler}. The propagation velocity is clearly different for the different cases. For reference we have inserted a line $v(t-t_1)$, where $t_1$ corresponds to the location of the first dip in each plot and the values $v$ are defined in the figure caption.

The table in Fig.~\ref{fig:FullModelResults}(g) shows the result of fitting a straight line to the location of the dip over time, that is the propagation velocity in units of $Ja/\hbar$ as well as the uncertainty of the fit. In general we see that for $L=24$, where we have more points, the uncertainty becomes smaller.For a more detailed discussion of the fits and uncertainties, see the supplemental material \cite{supplement}. It is difficult to say anything systematic about how the system size affects the velocity aside from this based on the limited number of results. The overall behavior is, however, consistent with the simple model of excitations. Specifically, $v_{{\rm SU}(2)},v_{\text{ABCD}}<4 Ja/\hbar$ while $v_{{\rm SU}(N)}$ approaches the SCBHM results with increasing $N$. For $N>2$, we see the expected enhancement when the initial state is entangled, while no enhancement is present for the ABCD product state consistent with the analytic model. This is a strong indicator that this model describes the enhancement mechanism. In the Mott limit where $v_{\rm SU(2)}=4Ja/\hbar$ and $v_{\rm SCBHM}=6Ja/\hbar$ are exact values for the maximum velocity within the doublon-holon model, the actual fitted velocities are slower as is also confirmed by a visual inspection of the lightcone.  Based on our analytic discussion, a natural conjecture for the general maximum velocity when starting from the Mott-insulating ground state is $v_{\text{SU}(N)}=\left(2K+2J\right)a/\hbar$, where $K$ is defined in Eq.~\eqref{eq:collectivehoppingK}. While the $(1-1/N)$ dependence is within the margins of error, achievable $L$ and $N$ are so limited that it is hard to say anything quantitative. In the SU($N$) case, it visually seems that propagation can be slightly faster than this simple limit, although the actual fitted velocities are smaller. The maximum velocities based on Eq.~\eqref{eq:collectivehoppingK} being slight underestimates for $N_{\rm P}>N$ is consistent with our numerical calculations in the supplemental material \cite{supplement} showing that $K$ tends to underestimate the spectrum bandwidth.

For bosonic and SU($N>2$)-superposition enhancement to occur, a relatively large value of $U_{\rm fin}/J$ is required and one expects the velocity to be dependent on $U_{\rm fin}/J$. For the SU(2) case and the ABCD product state on the other hand, we expect that the velocity should be less dependent on $U_{\rm fin}/J$.  Numerically it becomes more difficult to obtain good results as $U_{\rm fin}/J$ increases, so we are limited to smaller values. In Fig.~\ref{fig:FullModelResults}(g), we show $v_{{\rm SU}(N)}$ for $U_{\rm fin}/J\leq 12$.  For $U_{\rm fin}/J=0$ and $2$, all velocities are generally bounded by $4 J$. This means that for such a drastic quench the initial Mott insulator melts down almost completely so that the correlation is propagated by a single particle with the energy band $-2J \cos(k)$~\cite{Barmettler,Nagao-1}.  The cases of SCBHM and SU($N$) FHM with $N=3$ and $4$ display enhancement beyond this starting at $U_{\rm fin}/J=4$, although a value $U_{\rm fin}/J=6$ is required for behavior that is consistent with our expectations based on the simple model of excitations in the Mott limit.

{\it Summary and outlook.} We studied dynamics of the correlation propagation starting from a $1/N$-filled Mott insulator in the 1D SU($N$) FHM by means of both analytical and numerical calculations. For the SU(2) FHM, we showed that the collective enhancement of the effective hopping is impossible and the propagation velocity is bounded by $4Ja/\hbar$ for all values of the interaction, similar to the behavior of non-interacting bosons. For $N>2$, however, entanglement in the initial state gives rise to the collective enhancement of the hopping, leading to a higher velocity, and the velocity approaches 6$Ja/\hbar$, the same value as that of the SCBHM, at the large-$N$ limit.
We conjectured that for large lattice sizes the SU($N$) FHM will interpolate between the SU(2) FHM and SCBHM cases as $v_{\text{SU}(N)}=\left(4[1-1/N]+2\right)Ja/\hbar$. It will be interesting to examine this conjecture in future experiments. We numerically found that the collective enhancement is absent when the initial state is a simple product state. Since the high-temperature state of the flavor sector in the $1/N$-filled Mott insulator is a mixed state of all the permutations of the product state with equal probabilities, this strongly indicates that the enhancement of the velocity can only be observed for temperatures on the order of $T \simeq 0.1 J/k_{\rm B}$ or lower.

\section{Acknowledgments}
The calculations were performed utilizing the ITensor C++ library \cite{Fishman}. We acknowledge the support from Quantum Leap Flagship Program from MEXT [Grant No.~JPMXS0118069021], 
FOREST from Japan Science and Technology Agency (JST) [Grant No.~JPMJFR202T], 
and ASPIRE from JST [Grant No.~JPMJAP24C2]. MM would also like to acknowledge useful discussions on SU($N$) physics with Juan Polo.

\newpage

\clearpage
\onecolumngrid
\renewcommand{\thefigure}{S\arabic{figure}}
\renewcommand{\thetable}{S\arabic{table}}
\renewcommand{\theequation}{S\arabic{equation}}
\setcounter{equation}{0}
\setcounter{figure}{0}
\setcounter{table}{0}
\FloatBarrier
\begin{appendix}

\begin{center}
{
\large
Supplemental Material: ``Entanglement-enhanced correlation propagation in the one-dimensional SU($N$) Fermi-Hubbard model''
}
\end{center}

\section{S1. Restricted Fock-space representations for doublons and holons}

\subsection*{S.1. Doublon restricted Fock-space representations}
In this section we give a few more details on the representation of the Hilbert space restricted to one doublon in a system with $L=N_{\rm P}-1$ sites.

\subsubsection{S1.1.1. Determining the values $Q_1$,$Q_2$ and $K$}
In order to determine the values $Q_1$ and $Q_2$ which counts the number of doublon states that connect to only 1 neighboring state and two two neighboring states respectively, we use the binomial formula which counts the number of possible ways to pick $k$ objects from $n$ objects, which can also be used to count the number of ways one can distribute $k$ elements on $n$ sites.  To connect to two states we must first choose two flavors to represent a doublon on a given site, this can be done in $\binom{N}{2}$ ways. We must then choose one of the remaining flavors to occupy the adjacent site as this is what allows for connecting to two states, which can be done in  $\binom{N-2}{1}$ ways. We than need to distribute the remaining particles. Note that we have used one particle of 3 flavors, which means that these three flavors have $N_{\rm F}-1$ particles left that can be distributed on the remaining $L-2$ lattice sites in $\binom{L-2}{N_{\rm F}-1}\binom{L-N_{\rm F}-1}{N_{\rm F}-1} \binom{L-2 N_{\rm F}}{N_{\rm F}-1}$ ways. Once we have placed one set $N_{\rm F}-1$ there will be $N_{F}-1$ less lattice sites in which to replace the remaining, accounted for by the second and third term in in the product. Finally for $N>3$ there will also be $N-3$ sets of $N_{F}$ particles that must be distributed on the remaining lattice sites. Accounting for the particles removed with the placement of each flavor set this can be done in $\prod_{j=0}^{N-4} \binom{L+1-3N_{\rm F}-jN_{\rm F}}{N_{\rm F}}$ ways. Putting everything together we therefore get 
\begin{align}
Q_2 &= \binom{N}{2}\binom{N-2}{1}\binom{L-2}{N_{\rm F}-1}\binom{L-N_{\rm F}-1}{N_{\rm F}-1} \binom{L-2 N_{\rm F}}{N_{\rm F}-1} 
\prod_{j=0}^{N-4} \binom{L+1-3N_{\rm F}-jN_{\rm F}}{N_{\rm F}},
\end{align}
where the product in the second line has no contribution for $N=3$. 

The number of doublon states that only connects to one adjacent doublon state are counted in the following way. After we pick the two flavors representing the doublon in $\binom{N}{2}$ ways, we instead need to pick one of those two flavors for the adjacent site (blocking the movement of one of the particle flavors) which can be done in $\binom{2}{1}$ ways. We than have one flavor where we have removed two particles, one flavor where we have removed one particle and one flavor where we have removed 0 particles. Placing these sets in the remaining $L-2$ lattice sites can be done in $\binom{L-2}{N_{\rm F}-1}\binom{L-N_{\rm F}-1}{N_{\rm F}} \binom{L-2 N_{\rm F}-1}{N_{\rm F}-2}$ ways, where again we have accounted for having removed the number of lattice sites corresponding to the particles placed between each binomial coefficient. As before one finds that for $N>3$ there remains $N-3$ sets of $N_{\rm F}$ particles that can be placed in  $\prod_{j=0}^{N-4} \binom{L+1-3N_{\rm F}-jN_{\rm F})}{N_{\rm F}}$ ways. This leads to 
\begin{align}
Q_1&= \binom{N}{2}\binom{2}{1}\binom{L-2}{N_{\rm F}-1}\binom{L-N_{\rm F}-1}{N_{\rm F}} \binom{L-2 N_{\rm F}-1}{N_{\rm F}-2}  
\prod_{j=0}^{N-4} \binom{L+1-3N_{\rm F}-jN_{\rm F})}{N_{\rm F}},
\end{align}
where the product in the second line has no contribution for $N=3$.

Writing out the definition of the binomial coefficients and canceling out common terms the value $K$ is then given by Eq.~(4) of the main text.

In a similar way to find the total number of states, we simply pick out two flavors for the doublon in $\binom{N}{2}$ ways. Next we have to place the $N_{\rm F}-1$ particles of the two picked out flavors in the remaining $L-1$ sites. We then have to place the $N_{\rm F}$ particles of each remaining flavor in the remaining $L+1-2N_{\rm F}$ lattice sites. We can write this as 
\begin{align}
Q&=\binom{N}{2} \binom{L-1}{N_{\rm F}-1} \binom{L-N_{\rm F}}{N_{\rm F}-1}\prod_{j=0}^{N-3}\binom{L+1-(2+j)N_{\rm F}}{N_{\rm F}} \nonumber \\
&= \frac{N! (N_{\rm P}-2)!}{2(N-2)!(N_{\rm F}-1)!^2N_{\rm F}!^{N-2}},
\label{eq:localdoublondimension}
\end{align}
where the last line is obtained by writing out the definition of the binomial coefficients. 

\subsubsection{S.1.1.2. Large $N$-limit $N_{\rm P}=N$}
For the large $N$-limit of $N_{\rm P}=N$ ($N_{\rm F}=1$) the number of possible states that have one doublon on a given site simplifies to $Q=Q_2=\frac{N!}{2}$. We can think of all the possible combinations of having a doublon at site $j$ as belonging to a pseudo-lattice described by $L$ sites with local Hilbert spaces of dimension $Q$ label by $i_j\rangle$ where $j=0,...,Q-1$.  The local Hilbert-space can always be spanned by $Q$ vectors and using the linearly independent basis set $|i_j\rangle$ one can define a new linearly independent basis set in terms of the Fourier transform as
\begin{align}
| I_k \rangle =\frac{1}{\sqrt{Q}} \sum_{j=0}^{Q-1}e^{i \frac{2 \pi}{Q}j \cdot k}|i_j\rangle
\end{align} 
where $k=0,1,...,Q$. These are linearly independent and taking the inner product between the $k=0$ (corresponding to the equal superposition state $|d_i \rangle = \frac{1}{\sqrt{Q}}\sum_{j=0}^{Q} |i_j\rangle$) and any other state one obtains the well-known relation  
\begin{align}
\frac{1}{Q} \sum_{j=0}^{Q-1}e^{i \frac{2 \pi}{Q}j \cdot k} = \delta_{k,0}
\end{align} 

Let us next consider the matrix element of the Hamiltonian between the $k=0$ superposition state on one psuedosite and an arbitrary $k$-state on the adjacent psuedosite in terms of these these new basis states:
\begin{align}
\langle I_k |\hat{H}| (I+1)_{0} \rangle &=\frac{1}{Q}\sum_{j,j'=0}^{Q-1}e^{i \frac{2 \pi}{Q}j k} \langle i_{j'}| \hat{H} | (i+1)_{j}\rangle \nonumber \\
&= \frac{2}{Q}\sum_{j=0}^{Q-1}e^{i \frac{2 \pi}{Q}j k} = 2 \delta_{k,0}
\end{align}
Here we used the fact that one state always connects to two adjacent states with a value $J$ which means that the sum over $j'$ can be evaluated for any $j$ as $2J$ which can then be taken outside the sum. This shows that the $k=0$ local states never connect to any other $k \neq 0$ states under the action of the Hamiltonian and we can therefore solve the $k=0$ sector separately in terms of a simple non-interacting Hamiltonian given by 
\begin{equation}
\hat{H}_{\text{eff}} = -2J \sum_{j} \left(\hat{c}_{\text{eff},j}^\dagger \hat{c}_{\text{eff},j+1}+{\rm h.c.}\right)
\end{equation}
where $\hat{c}_{\text{eff},j}^\dagger$ creates the superposition states (and we are constrained to the one-particle sector $n_{\rm eff}=1$. This also holds for periodic boundary conditions in the original system in which case the eigenenergies are given by $-4J \cos(p)$ with $p=\frac{2 n \pi}{L}$ where $n=0,1...,L-1$.

\subsubsection{S.1.1.3. SU(2) case for general $N_{\rm P}$}
Finally we consider the case $N=2$ where $Q=Q_1=\frac{(N_{\rm P}-2)!}{(N_{\rm F}-1)!^2}$. We can think of this problem in two ways. Similar to before we can define this system in terms of a psuedo-lattice of size $L$ with a local Hilbert-space of dimension $Q$. Each state on site $j$ will connect to one state on site $j+1$ and for open boundary conditions the system Hamiltonian will actually split into exactly $Q$ different unconnected sectors each corresponding to a single-particle problem with an effective hopping $J$. Indeed numerical calculations in the restricted Fock-space also show that the spectrum will correspond to the single-particle spectrum for $J$ with a $Q$-degeneracy for each energy. We can consider these sectors as corresponding to different pseudoflavors. Periodic boundary conditions complicate the picture, however, as they will connect two different pseudoflavor sectors pairwise. For periodic boundary conditions, however, we can define the equal superposition state in the same way as we did for the $N=N_{\rm P}$ case. The pairwise connections between pseudoflavors then simply leads to a standard periodic boundary condition for the single-particle Hamiltonian that can be written in terms of the matrix-elements of these superposition states. The exact same argument using the basis of Fourier transforms leads to this being an unconnected sector and for periodic boundary conditions we expect that the groundstate spectrum is given by $-2J \cos(p)$ with $p=\frac{2 n \pi}{L}$ where $n=0,1,\ldots,L-1$.

\subsection*{S.1.2. Holon restricted Fock-space representations}
In this section we give a few more details on the representation of the $L=N_{\rm P}+1$ one holon restricted Hilbert-space. This is much simpler than the doublon-case and there are no specific cases which we have to consider separately. In this case the number of possible ways to have a holon on one site is as follows: After choosing the holon site there will be $\binom{L-1}{N_{\rm F}}$ ways to place one flavor $\binom{L-1-N_{\rm F}}{N_{\rm F}}$ ways to place the next flavor etc. which leads to 
\begin{align}
Q=\prod_{j=0}^{N-1} \binom{N_{\rm P}-j N_{\rm F}}{N_{\rm F}}=\frac{N_{\rm P}!}{(N_{\rm F}!)^N}.
\label{eq:localholondimension}
\end{align} 
We can  once again think of the system in terms of a pseudolattice with $L$ sites and local Hilbert space size $Q$. As for the SU(2) doublon case  each state on site $j$ will connect to one state on site $j+1$ and for open boundary conditions the system Hamiltonian will split into exactly $Q$ different unconnected sectors each corresponding to a single-particle problem with an effective hopping $J$. This is also corroborated by numerical ED calculations. Periodic boundary conditions will again introduce hopping between two pseudoflavors, but by taking the equal superposition state of all possible holons on one site $i$ a standard single particle Hamiltonian with periodic boundary conditions can be written in terms of their matrix elements. This sector will once again be disconnected by the same arguments as presented previously and we once again expect that the spectrum will be given by $-2J \cos(p)$ with $p=\frac{2 n \pi}{L}$ where $n=0,1,\ldots,L-1$. Again we emphasize that this is true for any $N$ and and $N_{\rm P}$ which means that holons will never have any speedup, similar to the bosonic case. 

\subsection*{S.1.3. ED calculations in the restricted Fock-spaces}

\begin{figure*}
\centering
\includegraphics[width=1.0\linewidth]{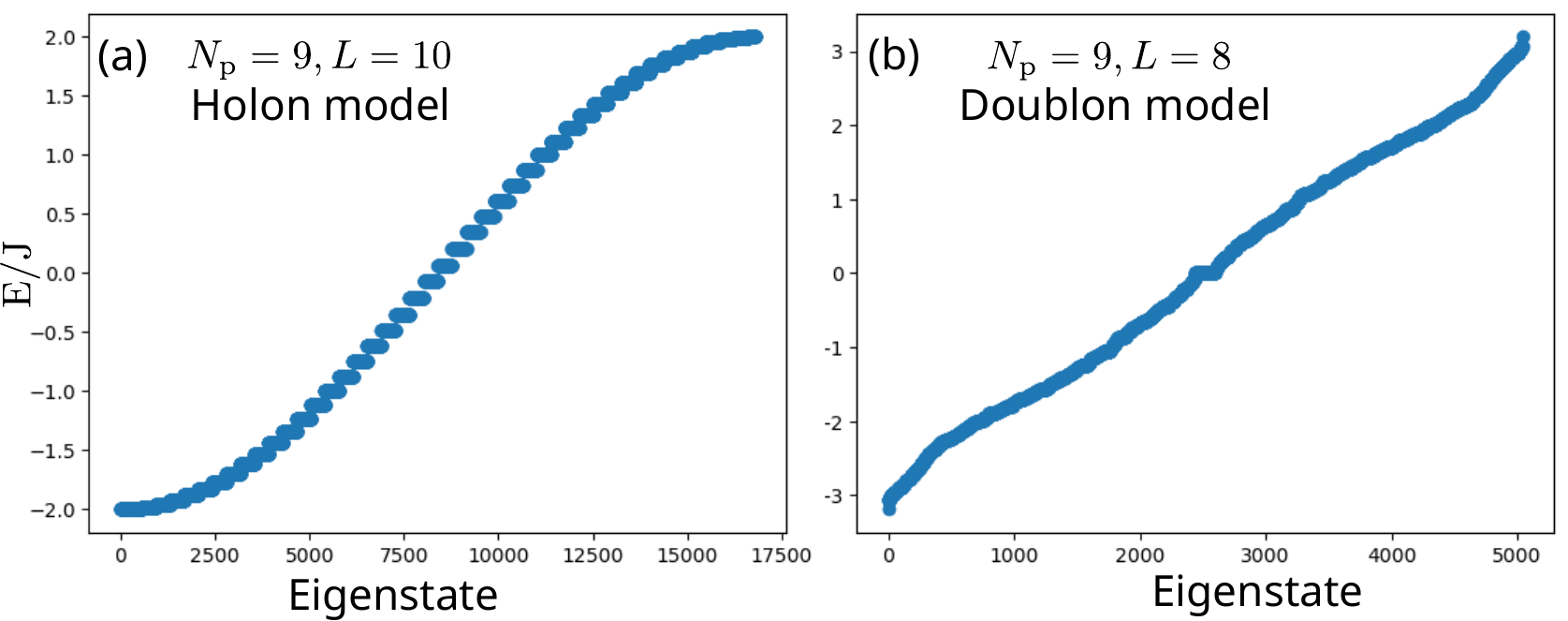}
\caption{The numerical spectrum for the restricted Fock spaces corresponding to the holon (a) and doublon (b) models for SU(3) with $N_{\rm P}=9$.}
\label{fig:holondoublonspectrum} 
\end{figure*}

Finally we show a few full exact-diagonalization (ED) calculations for the restricted Fock-space sectors. This is both to corroborate the above results and to investigate the doublon case for  $N>2, N_{\rm P}>N$, where we do not know the precise nature of the spectrum because we heuristically estimate the matrix element $K$ of the doublon hopping under the assumption of the equal superposition of each doublon configuration. Note that the dimension of the total Hilbert space corresponding to the restricted spaces grows exponentially with the the system size as $D=L Q$, where $Q$ is given by Eq.~(\ref{eq:localdoublondimension}) for the doublon model and Eq.~(\ref{eq:localholondimension}) for the holon model. The largest systems for which we can calculate the full spectrum is therefore limited to $N_{\rm P}=9$ for SU(3) and $N_{\rm P}=8$ for SU(4). This means that finite-size effects have a strong influence. 

In Fig.~\ref{fig:holondoublonspectrum}, we show the numerically obtained spectrum for the SU(3) $N_{\rm P}=9$ doublon ($L=8$) and holon ($L=10$) models. Note that these spectra has a large number of energies due to the large Hilbert-spaces, but the width of the spectrum $\Delta E = E_{\text{max}}-E_{\text{min}}$ is indeed determined by a separable sector with $-2J \cos(p)$ for the holon model. For the doublon model on the other hand, we do not know what the exact width should be as $N_{\rm P}>N$. It is however interesting to compare with a naive model which considers the $k=0$ equal superposition and a corresponding spectrum determined by $-2K \cos(p)$ with $p=\frac{2 n \pi}{L}$ where $n=0,1,\ldots,L-1$. We expect that this might give us a qualitative indicator of the behavior as $K$ contains the information about the ratio of states that connect to two adjacent sites vs one adjacent site. 

In Table~\ref{table:1}, we compare the numerical values of $\Delta E$, $K$ and analytic values $\Delta E_K$ corresponding to the naive model. For $N=N_{\rm P}$, the prediction of the naive model becomes exact as per the arguments in Sec.~S1.1.2 and this is borne out by the numerical results. For $N_{\rm P}>N$, the naive model consistently underestimates $\Delta E_K$. Nonetheless, qualitatively the naive model has a similar behavior to the numerical results. Particularly the relative values of $\Delta E_K$ between different values of $N_{\rm P}$ has the same ordering as for $\Delta E$.
\begin{table*}[]
\begin{tabular}{|l|l|l|l|l|l|}
\hline
 Model & SU(3) $N_{\rm P}=3$& SU(3) $N_{\rm P}=6$  & SU(3) $N_{\rm P}=9$  &  SU(4) $N_{\rm P}=4$ & $N_{\rm P}=8$ \\ \hline
$K/J$ & 2.000 & 1.50 & 1.429 & 2.00 &  1.857 \\ \hline
$\Delta E_{K}/J$ & 8.000& 5.427  & 5.714  & 6.000 & 6.337  \\ \hline
$\Delta E/J$ & 8.000 & 6.231  & 6.384  & 6.000 & 7.036  \\ \hline
\end{tabular}
\caption{Results for the effective doublon model corresponding to small SU(3) and SU(4) models. $K$ given by Eq.~(3) in the main text, $\Delta E_{K}$ based on $-2K \cos(p)$ with $p=\frac{2 n \pi}{L}$ where $n=0,1,\ldots,L-1$ and $\Delta E$ based on exact diagonalization are shown.}
\label{table:1}
\end{table*}

\section{S2. Generic equivalence between the ground-state properties of SU($N$) fermions and $N_{\rm P}$ bosons when $N\geq N_{\rm P}$}
In the previous section, we investigated the one doublon and one holon restricted excitation subspaces describing them in terms of an effective free Hubbard Hamiltonian and showed that the ground-state subesector of the SU($N$) free Fermi-Hubbard Hamiltonian was equivalent to the bosonic Hamiltonian when the number of flavors exceeds the total particle number. This is in fact a special case of a more general equivalence between the ground state sector of SU($N$) fermions and single-component bosons for $N\geq N_{\rm P}$ in the 1 particle per flavor (1PF) subsector. In this section we will show that the matrix elements of the SU($N$) symmetric $k$-reduced density matrix ($k$-RDM), i.e.,
\begin{align}
\hat{\rho}_{\text{SU}(N)}(l_1,...,l_{k},m_1,...,m_{k})=\sum_{\alpha_1,..\alpha_{k}=1}^N \prod_{j=1}^{k}\hat{c}_{l_j,\alpha_j}^\dagger \prod_{j=1}^{k} \hat{c}_{m_j,\alpha_j},
\label{eq:reduceddensityoperatorSUN}
\end{align}
will contain the matrix elements of the bosonic $k$-RDM as an embedded subsector corresponding to the states obeying SU($N$) symmetry in the 1PF sector. This means that any standard Hamiltonian which obeys SU($N$) symmetry such as the two-body Hamiltonian 
\begin{align}
H_{\text{SU}(N)}&= \sum_{\alpha} \sum_{jk} h_{jk} \hat{a}_{j,\alpha}^\dagger \hat{a}_{k,\alpha}+ 
\sum_{\alpha,\beta}\sum_{jklm} V_{jklm} \hat{a}_{j,\alpha}^\dagger \hat{a}_{k,\beta}^\dagger\hat{a}_{l,\beta} \hat{a}_{m,\alpha}.
\label{eq:genericbosonicHamiltonianSUN}
\end{align}
will similary have the bosonic Hamiltonian embbeded as a separable subsector corresponding to the states obeying SU($N$) symmetry in the 1PF sector. As the ground state of any Hamiltonian with SU($N$) symmetry obeys SU($N$) symmetry, the ground state therefore belongs to this bosonic subsector.

We consider a system with $N=N_{\rm P}$ fermionic flavors in $L$ sites/modes. In particular, by considering the one particle per flavor (1PF) sector, $N_1=N_2=...=N_N=1$, the problem can be reinterpreted as $N$ copies of a single fermion of each flavor occupying one of the $L$ sites. To establish the equivalence with bosons we are interested in mimicking the bosonic Fock-space. This can be done by considering all combinations of fermionic flavor Fock states that lead to a given total number of particles on each site. That is, we consider $|n_1,..., n_L \rangle_f$ where $n_k = \sum_{\alpha} n_{k,\alpha}$ (it is allowed for $n_k$ to be zero). Any such set contains numerous different Fock-space combinations denoted by the index $f$ (for example, a value of 2 on a site can be obtained by two particles AB, AC, BC, AD etc.). The number of states corresponding to a given sum of particles on each site can be found using elementary combinatorics. First, we chose $n_1$ flavors out of $N$. Next, we choose $n_2$ flavors out of $N-n_1$, etc. Each state within these sets is orthonormal, and we can define their spans as the local Hilbert spaces for that particular number combination, with a dimension given by the number of states in the combination 
\begin{align}
D_{n_1,...,n_L} =\frac{N!}{\prod_{j=1}^L n_j!}.
\end{align}
Another set of orthonormal basis states can be obtained by taking the Fourier transform of these states, \textit{i.e.}
\begin{align}
&|n_1,...,n_L \rangle_p = \nonumber \\
&\frac{1}{\sqrt{D_{n_1,...,n_L}}} \sum_{f=0}^{D_{n_1,...,n_L} -1}e^{i \frac{2 \pi}{D_{n_1,...,n_L}}f \cdot p}|n_1,..., n_L \rangle_f.
\label{eq:fourier_state}
\end{align}  
As in the previous section the $p=0$ state corresponds to a uniform superposition, now of all states with a given total particle number on each site, i.e. 
\begin{align}
|n_1,...,n_L \rangle_{0} =\frac{1}{\sqrt{D_{n_1,...,n_L}}} \sum_{f=0}^{D_{n_1,...,n_L} -1}|n_1,...,n_L \rangle_f.
\end{align}  
This uniform superposition state is invariant under the exchange of any two flavors and therefore obeys the SU($N$) symmetry.

Our goal is to show that the matrix elements of the bosonic $k$-RDM and hence standard number-preserving Hamiltonians are embedded as a disconnected symmetry sector in the collective $k$-RDM defined in Eq.(\ref{eq:reduceddensityoperatorSUN}), that is we need to evaluate
\begin{align}
\langle n'_1,...,n'_L |_0 \hat{\rho}_{\text{SU}(N)}(l_1,...,l_{k},m_1,...,m_{k}) |n_1,...,n_L \rangle_{p} \nonumber\\
\end{align}
in the 1PF sector.  To do this, we insert the definition of the states of Eq.\ (\ref{eq:fourier_state}), which leads to 
\begin{align}
&\langle n'_1,...,n'_L |_0  \hat{\rho}_{\text{SU}(N)}(l_1,...,l_n,L_1,...,m_n)  |n_1,...,n_L \rangle_p = \nonumber \\
& \frac{1}{\sqrt{D_{n'_1,...,n'_L}}\sqrt{D_{n_1,...,n_L}}}  \sum_{f=0}^{D_{n_1,...,n_L} -1} e^{i \frac{2 \pi}{D_{n_1,...,n_L}}f \cdot p} \times \nonumber \\ 
&\sum_{f'=0}^{D_{n'_1,...,n'_L} -1}\langle n'_1, ...,n'_L |_{f'} \hat{\rho}_{\text{SU}(N)}(l_1,...,l_n,m_1,...,m_n) |n_1,...,n_L \rangle_f.
\end{align}
We consider the case where an annihilation or creation operator is never used on the same site more than once first. We will explain how allowing for multiple creation or annihilation operators on a single site would change the result afterward. To proceed, we must evaluate the matrix elements inside the sum.

Non-zero matrix elements are unambiguously only obtained when $n'_{l_j}=n_{l_j}+1$, $n'_{m_j}=n_{m_j}-1$ and $n'_{q}=n_{q}$ where $q \neq m_j,l_j$ for all $j$. Any other combination would give zero due to the orthonormality of the Fock states.  Additionally, we see that an annihilation operator of a given flavor is always paired with a creation operator of the same flavor. For any Fock state which has $n_{m_j}$ total number of particles on the site $m_j$, there are exactly $n_{m_j}$ flavors represented on the site and the same flavor cannot be present on any other site due to the 1PF sector restriction. Any of these pairs of creation and annihilation operators of the same flavor therefore leads to exactly $n_{m_j}$ candidates for non-zero matrix-elements, each with a value of 1. From this, the sum over $f'$ for any matrix-element is given by $\prod_{j=1}^{k} n_{m_j}$. These can then be taken outside the $f$ sum since this value is solely determined by the total number of particles on each site, which is always the same for a given choice of $|n_1,..., n_L \rangle$ and the specific indices of the $k$-RDM. This allows us to evaluate the sum as 
\begin{align}
\sum_{f=0}^{D_{n_1,...,n_L} -1} e^{i \frac{2 \pi}{D_{n_1,...,n_L}}f \cdot p}= D_{n_1,...,n_L} \delta_{p,0}.
\end{align}
The matrix elements can therefore be written as 
\begin{align}
&\langle n'_1,...,n'_L |_0  \hat{\rho}_{\text{SU}(N)}(l_1,...,l_{k},m_1,...,m_{k})|n_1,...,n_L \rangle_p \nonumber \\
&= \frac{\prod_{j=1}^{k} n_{m_j} \delta_{n'_{m_j},n_{m_j-1}}\delta_{n'_{l_j},n_{l_j}+1}  \prod_{q \neq m_j,l_j}\delta_{n_q,n_q'}}{\sqrt{D_{n'_1,...,n'_L}}\sqrt{D_{n_1,...,n_L}}} D_{n_1,...,n_L}\delta_{p,0} \nonumber \\
&=\frac{\prod_{j=1}^{k} n_{m_j} \frac{N!}{\prod_{q=1}^L n_q!}}{\sqrt{\frac{N!}{\prod_{q=1}^L n'_q!}\frac{N!}{\prod_{q=1}^L n_q!}}} \delta_{p,0}  \prod_{j=1}^{k}\delta_{n'_{m_j},n_{m_j-1}}\delta_{n'_{l_j},n_{l_j}+1}  \prod_{q \neq m_j,l_j}\delta_{n_q,n_q'} \nonumber \\
& =\prod_{j=1}^{k} n_{m_j} \sqrt{\frac{\prod_{q=1}^L n'_q!}{\prod_{q=1}^L n_q!}} \delta_{p,0}  \prod_{j=1}^k\delta_{n'_{m_j},n_{m_j-1}}\delta_{n'_{l_j},n_{l_j}+1}  \prod_{q \neq m_j,l_j}\delta_{n_q,n_q'}. 
\end{align}
Here we know that $n_q'=n_q$ for any index $q\neq m_j,l_j$, which means that these terms cancel out in the fraction. This means that only terms connected by the $k$-RDM remain in the fraction, and they can be evaluated as 
\begin{align}
&\langle n'_1,...,n'_L |_0  \hat{\rho}_{\text{SU}(N)}(l_1,...,l_k,m_1,...,m_k)|n_1,...,n_L \rangle_p \nonumber \\
&= \prod_{j=1}^{k} n_{m_j} \sqrt{\prod_{j=1}^{k}\frac{(n_{m_j}-1)!(n_{l_j}+1)!}{n_{m_j}! n_{l_j}!}}\delta_{p,0} 
 \prod_{j=1}^{k} \delta_{n'_{m_j},n_{m_j-1}}\delta_{n'_{l_j},n_{l_j}+1}  \prod_{q \neq m_j,l_j}\delta_{n_q,n_q'}   \nonumber \\
&=  \prod_{j=1}^{k} \sqrt{n_{m_j}} \sqrt{n_{l_j}+1} \delta_{p,0}  \prod_{j=1}^{k}\delta_{n'_{m_j},n_{m_j-1}}\delta_{n'_{l_j},n_{l_j}+1}  \prod_{q \neq m_j,l_j}\delta_{n_q,n_q'}.
\label{eq:product_k_rdm_simple}
\end{align}
This only holds if there are no multiple applications of creation/annihilation operators on the same site otherwise the result is trivially modified. If there are $k'$ annihilation and/or $k''$ creation operators used on the same sites $s$ and/or $r$, respectively, the non-zero elements are now given by the restriction $n'_{r}=n_{r}+k''$, $n'_{s}=n_{s}-k'$, leading to a similar update of the Kronecker delta. Additionally, once the annihilation operator has been applied once to a given site $s$ the created $n_s$ states each have the new occupation value $n_s-1$, and this updated value must be used when the $n_s-1$ states are created by the next application of the annihilation operator. In this way, the expectation value is modified as $\prod_{j=0}^{k'-1}(n_{s}-j) \prod_{j,m_j\neq s}n_{m_j}$. Overall, this leads to the following modification of Eq.\ (\ref{eq:product_k_rdm_simple}):
\begin{align}
&\langle n'_1,...,n'_L |_0  \hat{\rho}_{\text{SU(N)}}(l_1,...,l_k,m_1,...,m_k)|n_1,...,n_L \rangle_p \nonumber \\
&= \prod_{j=0}^{k'-1}(n_{s}-j) \prod_{j,m_j\neq s}n_{m_j} \sqrt{\frac{\prod_{q=1}^L n'_q!}{\prod_{q=1}^L n_q!}} \delta_{p,0}  \delta_{n'_{s},n_{s}-k'}\delta_{n'_{r},n_{r}+k''} \prod_{j,m_j\neq s,r}\delta_{n'_{m_j},n_{m_j-1}}\delta_{n'_{l_j},n_{l_j}+1}  \prod_{q \neq m_j,l_j,s,r}\delta_{n_q,n_q'}  \nonumber \\
&= \prod_{j=0}^{k'-1}(n_{s}-j) \prod_{j,m_j\neq s}n_{m_j}  \sqrt{ \frac{(n_{s}-k')!(n_{r}+k'')!}{n_{s}! n_{r}!} \prod_{j,m_j \neq s,r} \frac{(n_{m_j}-1)!(n_{l_j}+1)!}{n_{m_j}! n_{l_j}!}} \times \nonumber \\
&\delta_{p,0}  \delta_{n'_{s},n_{s}-k'}\delta_{n'_{r},n_{r}+k''} \prod_{j,m_j\neq s,r}\delta_{n'_{m_j},n_{m_j-1}}\delta_{n'_{l_j},n_{l_j}+1}  \prod_{q \neq m_j,l_j,s,r}\delta_{n_q,n_q'}   \nonumber \\
&= \prod_{j=0}^{k'-1} \sqrt{n_{s}-j} \prod_{j=1}^{k''} \sqrt{n_{r}+j}  \prod_{j,m_j \neq s,r} \sqrt{n_{m_j}} \sqrt{n_{l_j}+1} \delta_{p,0}  \delta_{n'_{s},n_{s}-k'}\delta_{n'_{r},n_{r}+k''} \prod_{j,m_j\neq s,r}\delta_{n'_{m_j},n_{m_j-1}}\delta_{n'_{l_j},n_{l_j}+1}  \prod_{q \neq m_j,l_j,s,r}\delta_{n_q,n_q'} . 
\end{align}

If multiple sites have the application of more than one creation or annihilation operator, the same modification has to be made for each of these. In all cases, the $p=0$ Fourier state leads to the same matrix-elements, obtained from the equivalent bosonic $k$-RDM in the standard Bosonic Fock-space $ |n_1,..., n_L \rangle$ and  $ \langle n'_1,..., n'_L |$. They obey the SU($N$) symmetry due to being a uniform superposition of all combinations.  Additionally, we see that the $k=0$ sector forms a disconnected symmetry sector and any SU($N$) fermionic Hamiltonian will therefore have the same energies in this symmetry sector as the single-component bosonic Hamiltonian. 

\section{S3. Numerical representation of SU($N$) utilizing matrix product states and tensor networks} 
In order to numerically investigate the  SU($N$) Fermi-Hubbard model we utilize the ITensor library (See Ref.~[48] in the main text). Within the standard library  the representation of spinless fermions and spinful electrons on a lattice is already efficiently implemented. We therefore build on top of this and represent the $N$-component Hubbard model in terms of these building blocks. For even $N$ we utilize the two-component electron representation build into the library with distinct sublattices corresponding to different flavors. For example the SU(6)-model is represented by the one-dimensional Hamiltonian
\begin{align}
\hat{H} &= -J \sum_{j=1}^{L-1} \hat{c}_{\uparrow,j}^\dagger \hat{c}_{\uparrow,j+3} + \hat{c}_{\downarrow,j}^\dagger \hat{c}_{\downarrow,j+3}  + {\rm h.c.} 
+U \left(\sum_{\mu,\mu'}\sum_{j=1,4,\ldots}^{3 L-2} \hat{n}_{\mu,j} \hat{n}_{\mu',j+1}+\hat{n}_{\mu,j} \hat{n}_{\mu',j+2}\right)
+U\sum_{j,\mu \neq \mu'}  \hat{n}_{\mu,j} \hat{n}_{\mu',j}
\end{align} 
where $\mu,\mu'=\uparrow,\downarrow$. If we name the sublattices $A,B,C$ the flavors correspond to $A_{\uparrow},A_{\downarrow},B_{\uparrow},B_{\downarrow},C_{\uparrow},C_{\downarrow}$. The SU(4)-model can be represented in a similar way, but is simpler as only two sublattices are required. 
  
For the SU(3) calculations we utilize a system of spinless fermions with 3 sublattices that is the Hamiltonian is given by 
\begin{align}
\hat{H} &= -J \sum_{j} \left(\hat{c}_{j}^\dagger \hat{c}_{j+3}  + {\rm h.c.} \right)
+U \left(\sum_{j=1,4,\ldots}^{3 L-2}  \hat{n}_{j} \hat{n}_{j+1}+\hat{n}_{j} \hat{n}_{j+2}\right)
\end{align} 
with the flavors corresponding to the $A,B,C$ sublattices.

In order to efficiently apply the Suzuki-Trotter decomposition to the MPS representation  of the wavefunction as a series of gates $T_{j,j+1}$ we use use fermionic swap-gates to exchange neighboring sites.

\section{S4. Velocity fits} 
\begin{figure*}
\centering
\includegraphics[width=1.0\linewidth]{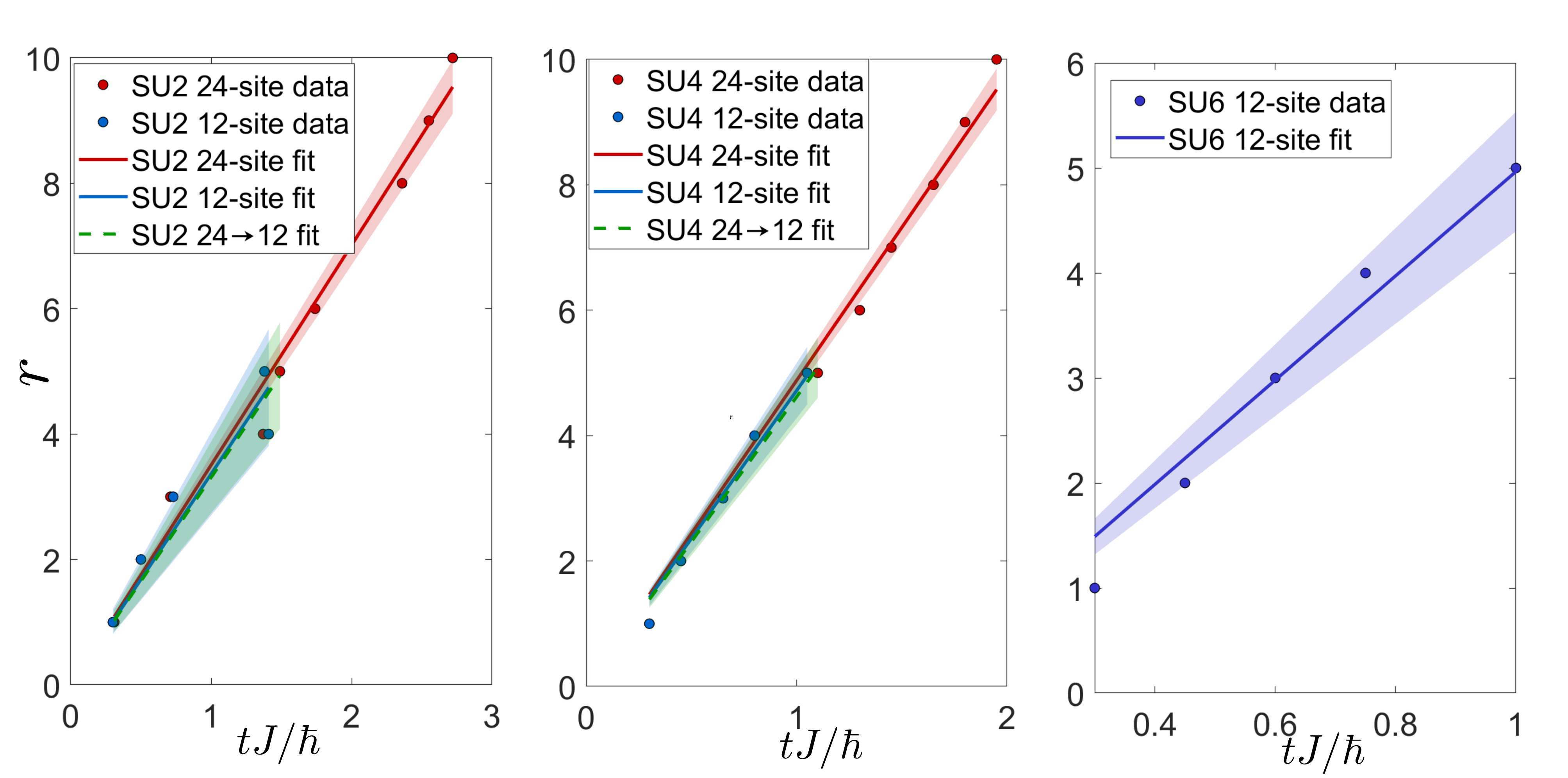}
\caption{Distance as a function of the time at which the first dip in the correlation function is observed at that distance. These data in Figs.~\ref{fig:velocityfits}(a), (b), and (c) are extracted from the correlation-spreading plots in Figs.~2(a), (e), and (c) in the main text. For SU(2) and SU(4), minima based on the simulation data for both $L=12$ and $L=24$ is shown. The lines correspond to fits based on the data, with the striped line corresponding to a fit based on the $L=24$ simulation data restricted to the first five points. The shaded areas are based on the 95\% confidence intervals of the fitted velocity.  }
\label{fig:velocityfits} 
\end{figure*}

In this section, we show a few examples of the velocity fits used in the main paper. We investigate SU($2$), SU($4$) and SU($6$) when an initial ground state at $U_{\rm ini}=40J$ is quenched by a Hamiltonian with $U_{\rm fin}=8J$. In the main text, fits based on $L=12$ simulation data had quite large uncertainties; in this section we investigate the source of these comparing with $L=24$ simulation data. We therefore compare with two fits based on the $L=24$ simulation data, one using all available distances and one restricted to the distances achievable for the $L=12$ simulation. The fitted coefficients can be found in Table \ref{tab:velocityfits} while plots of the raw data versus fits can be found in Fig.~\ref{fig:velocityfits}. 

\begin{table}

\begin{tabular}{|c|c|c|c|}
 \hline
          & $L=12$ &$L=24$ restricted to $L=12$ & $L=24$   \\
    \hline
    SU($2$) & $3.36 \pm0.66$ & $3.31 \pm0.57$ & $3.51 \pm0.16$ \\
    \hline
   	SU($4$) & $4.72 \pm0.44$  & $4.62 \pm0.44$ & $4.88 \pm 0.17$  \\
   	  \hline
   	SU($4$) $r=1$ removed & $4.77 \pm0.31$  & $4.67 \pm0.37$ & $4.89 \pm0.16$  \\
    \hline
    SU($6$) & $4.97 \pm0.57$  & - & - \\
    \hline
     SU($6$) $r=1$ removed & $5.04 \pm0.44$  & - & - \\
    \hline
\end{tabular}
\caption{The fitted velocities and their uncertainty. The second column is based on the $L=24$ simulation data but with a fit restricted to the same number of distances as the $L=12$ simulation for comparison.}
\label{tab:velocityfits}
\end{table}

The largest uncertainty for $L=12$ is in the SU($2$) fit. Looking at the plot, this is because of a jump in the peak location between the first three $r$'s ($r=1,2,3$) and the next two $r$'s ($r=4,5$), which is also clearly visible in Fig.~2(a) in the main text and heavily influences the five-point linear fit. This jump is also present for the $L=24$ simulation so it is not an artefact of the small system size per se, and the uncertainty based on fitting only the first five points for the latter is not much improved. However, since the next five points behave monotonously, the ten-point linear fit gives a more accurate velocity. For SU($4$) and SU($6$), the first point is a slight outlier which matters more when only the first five points can be used in the fit, less so when all the ten points can.  In principle smaller uncertainties can be obtained by excluding the first three points for SU($2$) (which only makes sense for $L=24$) and the first point from SU($4$) and SU($6$) as can be seen for the latter two in Table \ref{tab:velocityfits}. However, the $L=24$ case still has roughly half the uncertainty due to better statistics from the larger number of points. For SU($6$), where only the $L=12$ data is accessible removing the first data point, is arguably a good idea, but the justification is not entirely clear. We therefore use the fit based on all 5 points with the larger uncertainty in the main text. This emphasizes that results for the SU($6$) simulations are not as precise as in the other systems. Note that even removing the first point, the overall uncertainty is still larger than the SU($4$) case.    
\end{appendix}
\end{document}